\documentclass{acm_proc_article-sp}
\usepackage{graphicx}

\begin{document}

\conferenceinfo{}{Bloomberg Data for Good Exchange 2017, NY, USA}

\title{Machine Learning for Drug Overdose Surveillance}

\numberofauthors{2} 
\author{
\alignauthor Daniel B. Neill \\
  \affaddr{Event and Pattern Detection Laboratory}\\
  \affaddr{Carnegie Mellon University}\\
  \affaddr{Pittsburgh, PA}\\
  \email{neill@cs.cmu.edu}
\alignauthor William Herlands \\
   \affaddr{Event and Pattern Detection Laboratory}\\
   \affaddr{Carnegie Mellon University}\\
   \affaddr{Pittsburgh, PA}\\
   \email{herlands@cmu.edu}
}

\maketitle 
%

\section{Introduction}
\label{sec:introduction}

A recent epidemic of opioid overdoses has garnered national attention in 
the United States~\cite{hhsopioid}. The CDC estimates that 47,055 
overdose deaths occurred in the U.S. in 2014, 61\% of which involved 
opioids (including heroin, pain relievers such as oxycodone, and 
synthetics)~\cite{cdc,neill2017overdose}. These statistics motivate 
public health agencies to identify emerging trends in overdoses so that 
they can better target prevention and response measures.  However, 
reliable identification of such trends requires us to solve several 
challenges.  First, to accurately pinpoint the affected spatial area, we 
must account for the complex, correlated nature of spatio-temporal data, 
modeling these correlations and using them to distinguish between 
significant anomalous patterns and fluctuations due to correlated noise.  
Second, when a pattern of overdoses differentially affects some 
subpopulation (such as elderly males, or drug users who combine heroin 
with alcohol), it is important to identify and precisely target that 
subpopulation.  To do so, we must scan over multiple dimensions of the 
data including demographics and behaviors as well as geographic areas. 
We expect that this will also increase power to detect emerging overdose 
clusters in their early stages, when case counts are low.  Most importantly, we must consider how to integrate information across 
space, time, and subpopulation-level characteristics, since an event of 
interest may be spread across all of these features.  Though individual 
anomaly detection methods are commonly used in 
practice~\cite{kowalska2012maritime,stegle2008gaussian}, simply 
considering the anomalousness of each individual point loses power to 
detect subtle or emerging anomalies.

\section{Methods}

We consider two recently proposed extensions of the fast subset scan~\cite{neill2012fast}, which detects anomalous patterns by efficiently maximizing a log-likelihood ratio statistic over subsets of data points.  These approaches, the Gaussian Process Subset Scan (GPSS) and Multidimensional Tensor Scan (MDTS), are described in more detail in~\cite{herlands2017} and~\cite{neill2017overdose} respectively.  For each method, we frame the search as a log-likelihood ratio (LLR) comparison between a null hypothesis $H_0$, assuming no events of interest, and a set of alternative hypotheses $H_1(S)$, each representing the occurrence of an event in some subset $S$.
GPSS monitors data $D = (x,y)$, where $x = \{x_1,\dots,x_n\}$, $x_i \in \mathbb{R}^D$ are covariates, and $y = \{y_1,\dots,y_n\}$, $y_i \in \mathbb{R}$ is a response variable.  For example, each $x_i$ could represent the coordinates of a location in space and time, and the corresponding $y_i$ could represent the number of overdose deaths for that location and time period.  In MDTS,
we again monitor data $D = (x,y)$, but now each $x_i$ represents a set of discrete-valued characteristics.  For example, each $x_i$ could represent the characteristics of a given overdose victim or set of identical victims (gender = male, age group = 20-29, etc.) and the corresponding $y_i$ would be the number of individuals with those characteristics (i.e., $y_i = 1$ for a single case).  Both approaches follow the same general framework:
\begin{enumerate}
    \item \textbf{Estimate the distribution of the observed responses $y$ under the null hypothesis $H_0$.}
    
    \noindent GPSS assumes that points are drawn from a function with a Gaussian process prior, $y = f(x) + \epsilon$, where $f(x) \sim GP(m(x), k(x,x'))$ and $\epsilon \sim \mathcal{N}(0,\sigma_\epsilon^2 I)$.  Parameters of the Gaussian process are learned from the entire dataset $D$. For a given subset $S$, the $y_i$ will follow a multivariate Gaussian distribution, and we condition on $D \setminus S$ to infer the posterior mean vector $\mu$ and covariance matrix $\Sigma$.
    
    Similarly, MDTS performs scalable tensor decomposition to model $D$ as a sum of rank-one tensors. For a given subset $S$, MDTS uses the learned tensor decomposition to estimate the mean vector $\mu$, and assumes that each $y_i$ is drawn independently from a Poisson distribution with mean $\mu_i$.
    \item \textbf{Define a search space of subsets $S$ to consider.}  
    
    GPSS forms the local $k$-neighborhoods $S_{ik}$ defined by each $x_i$ and its $k-1$ nearest neighbors according to some distance metric, then considers all $S \subseteq S_{ik}$.  In other words, GPSS conducts an unconstrained search within each neighborhood to find the subset which maximizes the LLR. This provides GPSS with the flexibility to identify highly irregular shapes which still satisfy the spatial proximity constraint.
    
    MDTS considers the subspaces formed by Cartesian products of subsets of values for each attribute.  For example, given attributes gender and age, one valid subspace would be ``males and females in age groups 10-19 and 20-29'', while the subset of ``males aged 10-19 and females aged 20-29'' would not be considered.
    
    \item \textbf{Define the alternative hypothesis $H_1(S)$.}  
    
    GPSS assumes that under $H_1(S)$, the values of $y_i$ for $x_i \in S$ are drawn from a multivariate Gaussian with a mean shift of $\beta$: $y \sim \mathcal{N}(\mu + \beta w, \Sigma)$, where $w$ is a binary vector such that $w_i = 1\{x_i \in S\}$.  
    
    MDTS assumes that under $H_1(S)$, the values of $y_i$ for $x_i \in S$ are increased by a multiplicative factor of $q > 1$, i.e., $y_i \sim \mbox{Poisson}(q\mu_i)$ for $x_i \in S$ and $y_i \sim \mbox{Poisson}(\mu_i)$ for $x_i \not\in S$.
    
    \item \textbf{Maximize the log-likelihood ratio over subsets.} 
    
    For both methods, the LLR score of a given subset $S$ is defined as $F(S) = \log \frac{Pr(Data \:|\: H_1(S))}{Pr(Data \:|\: H_0)}$.  The subsets $S^\ast = \arg\max_S F(S)$ are identified as the most anomalous subsets, and randomization testing is performed to determine a threshold for statistical significance.
    
    To avoid the exponential time complexity of searching over all $2^k$ subsets for a neighborhood of size $k$, GPSS uses a novel approximation technique which extends~\cite{speakman2016penalized} to iteratively compute conditionally optimal subsets.  More details are provided in~\cite{herlands2017}.
    
    Similarly, to avoid exhaustively searching over all $2^V$ subspaces, where $V$ is the sum of the arities of all discrete-valued attributes in the data, MDTS uses an iterative conditional optimization technique which uses the linear-time subset scanning property~\cite{neill2012fast} to efficiently and exactly compute the conditionally optimal subset of values for a given attribute, given the current subsets of values for all other attributes.  This conditional optimization step is repeated until a local optimum is reached, and multiple restarts are used to approach the global optimum.  More details are provided in~\cite{neill2017overdose}. 
\end{enumerate}

GPSS and MDTS have multiple, complementary advantages.  GPSS is useful for modeling and accounting for non-iid correlations in real-world systems, such as spatio-temporal urban data.  GPs provide a natural means of learning covariance structure from data, extend to multiple dimensions, and seamlessly handle missing data. Given case data, MDTS has high power to detect and characterize emerging trends which may only affect a subset of the monitored population (e.g., specific ages, genders, neighborhoods, or users of particular drugs or combinations of drugs).

\section{Results}

We now describe retrospective case studies applying our surveillance 
techniques to real-world overdose data from two different areas of the country.  
First, we use the Gaussian Process Subset Scan to detect and localize 
overdose clusters in aggregated spatio-temporal data from New York City.  
Second, we apply the Multidimensional Tensor Scan to identify 
subpopulation-level overdose trends in case data from western Pennsylvania, 
and explore the discovered clusters in collaboration with county public 
health officials.

\subsection{Case study 1: aggregated count data}

\begin{figure*}[t]
\begin{center}
\includegraphics[width=7cm]{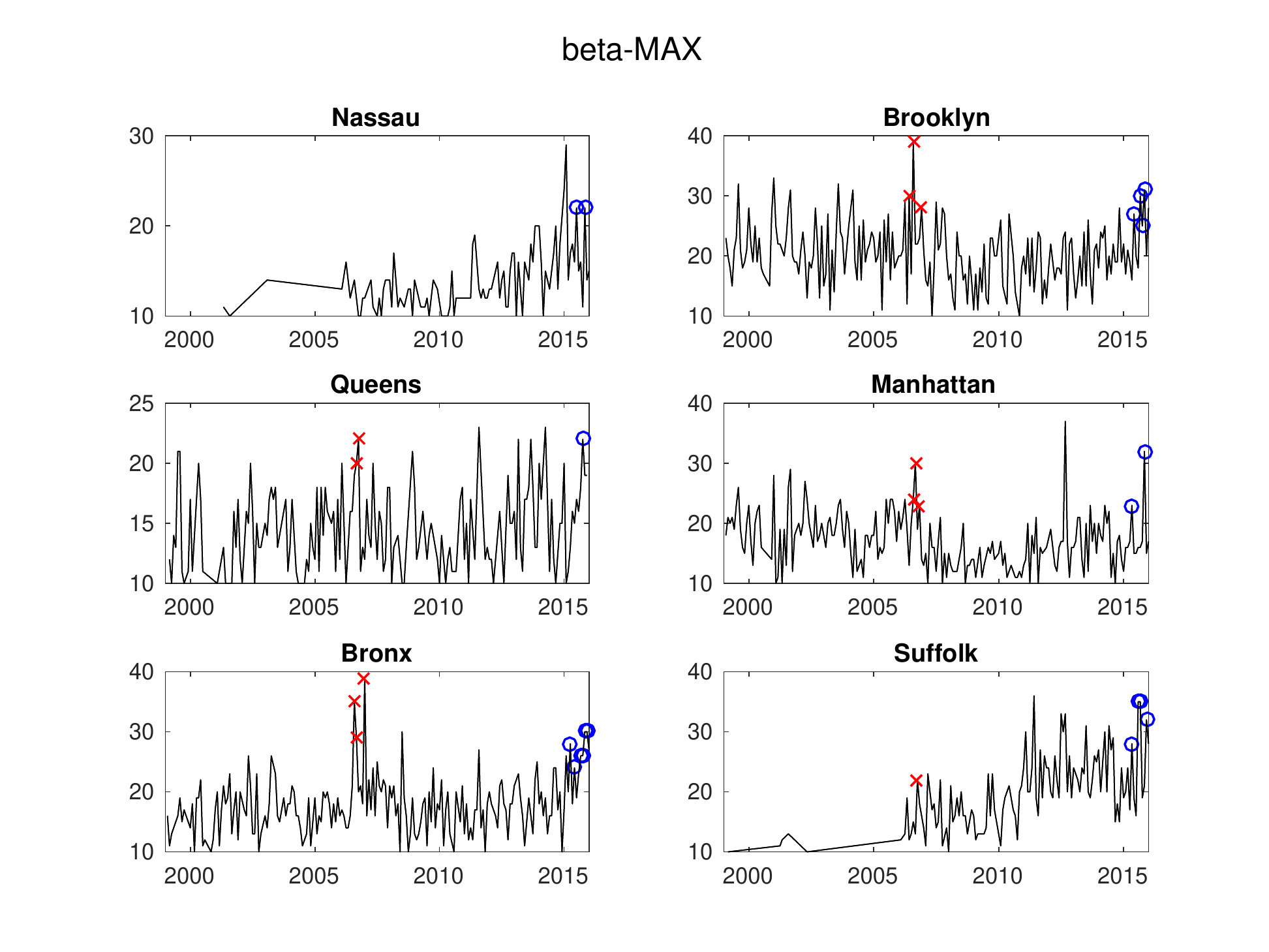}
\includegraphics[width=7cm]{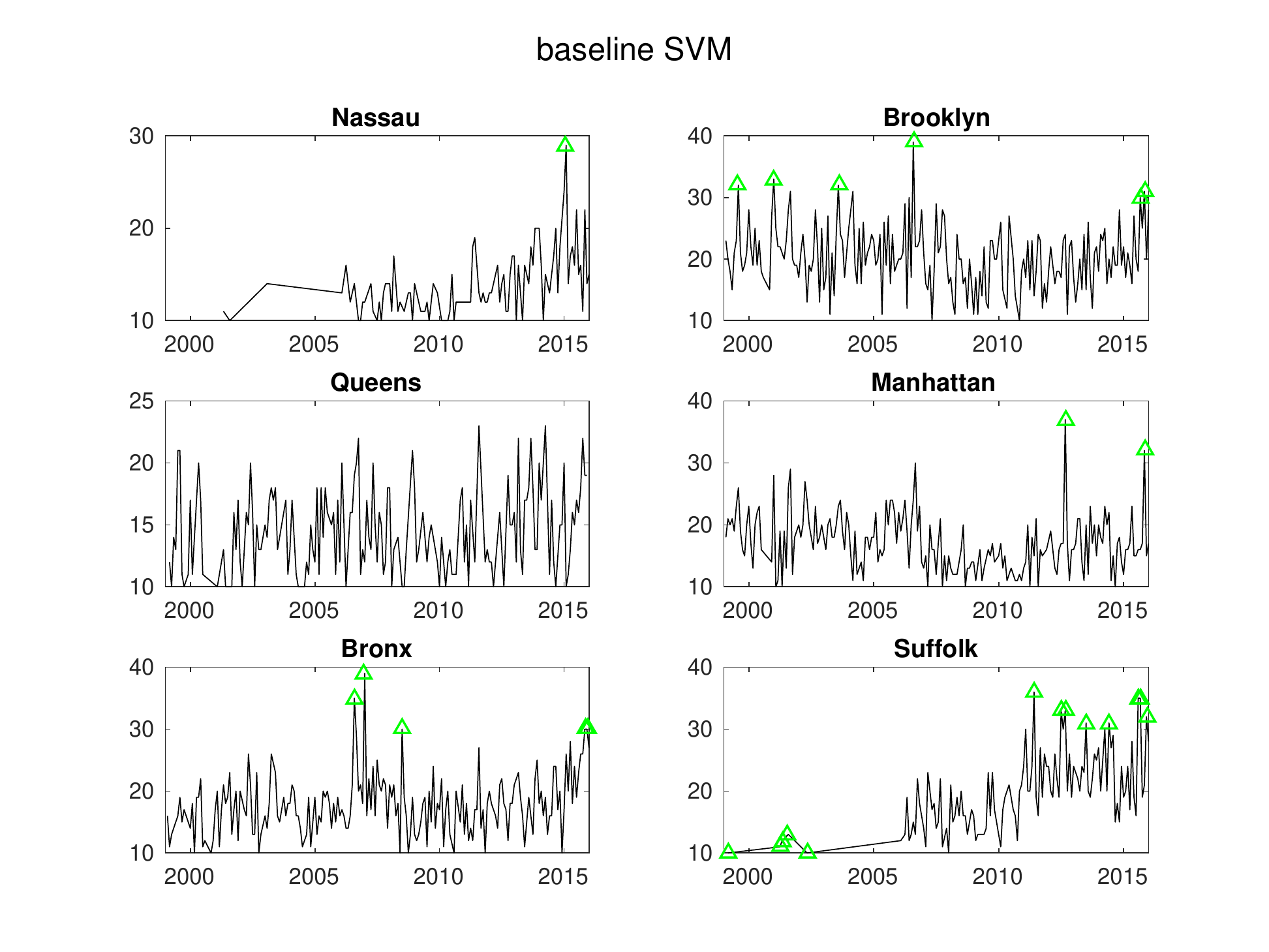}
\caption{Monthly opioid overdose deaths in the NYC metropolitan area from 1999-2015.  In the left panel, the two statistically significant anomalies detected by GPSS are depicted as red crosses and blue circles.  The right panel shows significant anomalies detected by the one-class SVM.}
\label{fig:opioid}
\end{center}
\end{figure*}

For our first case study~\cite{herlands2017}, we analyzed monthly opioid overdose deaths in the New York City metropolitan area from 1999-2015~\cite{CDCWONDER}. Data are provided at a 
county level and include Manhattan, Brooklyn, Queens, the Bronx, Nassau County, and Suffolk County. Data records are missing for some months in some counties.  We compare GPSS against three competitive baseline algorithms, including GP anomaly detection~\cite{kowalska2012maritime,stegle2008gaussian}, one-class 
SVM~\cite{scholkopf2001estimating}, and robust multivariate outlier detection using the Mahalanobis distance~\cite{rousseeuw2005robust,rousseeuw1990unmasking}. We 
apply the GPSS and baseline approaches jointly to data across all locations and time steps, performing randomization testing at $\alpha=.05$ to identify significant clusters.  GPSS identifies two statistically significant anomalous patterns in the data, shown in blue circles and red crosses in 
Figure~\ref{fig:opioid} (left panel). The baseline anomaly detection methods (Figure~\ref{fig:opioid}, right panel) failed to discover a coherent anomalous pattern, instead selecting individual points across space and 
time.

The anomalies detected by GPSS correspond to important public health events. The blue circles at the end of 2015 indicate a surge in opioid deaths corresponding to a 
well known plague of fentanyl-related deaths in NYC~\cite{HealingNYC}. The anomaly denoted by red crosses in 2006 is particularly interesting since it indicates a 
spike in opioid deaths immediately preceding the introduction of community training programs to administer a lifesaving naloxone drug. This may indicate a surge in 
fatalities that was cut short by making naloxone more widely available and educating communities in its use.

\subsection{Case study 2: victim-level case data}

For our second case study~\cite{neill2017overdose}, we used the MDTS approach described above to analyze a publicly available dataset from the Allegheny County, PA 
medical examiner's office and to detect emerging overdose patterns and trends.  The data consists of approximately 2000 fatal accidental drug overdoses between 2008 
and 2015.  For each overdose victim, we have date, location (zip code), age decile, gender, race, and the presence or absence of 27 commonly abused drugs in their 
system.  The highest-scoring clusters discovered by MDTS were shared with Allegheny County's Department of Human Services and their feedback obtained.

One set of potentially relevant findings from our analysis involved fentanyl, a dangerous and potent opioid which has been a serious problem in western PA.  In 
addition to identifying two well-known, large clusters of overdoses (14 deaths in Jan 2014 and 26 deaths in Mar-Apr 2015), MDTS was able to provide additional, 
valuable information about each cluster.  For example, the first cluster was likely due to fentanyl-laced heroin, while the second was more likely due to fentanyl 
disguised as heroin (only 11 of the 26 victims had heroin in their system).  Moreover, the second cluster was initially confined to the Pittsburgh suburb of McKeesport 
and the typical overdose demographic of white males ages 20-49, before spreading across the county.  Our analysis demonstrated that prospective surveillance using MDTS
would have identified the cluster as early as March 29th, enabling more targeted prevention efforts.  

MDTS also discovered a previously unidentified, highly localized 
cluster of fentanyl-related overdoses affecting an unusual and underserved demographic (elderly black males near downtown Pittsburgh).  This cluster occurred in 
Jan-Feb 2015, and may have been related to the larger cluster of fentanyl-related overdoses that occurred two months later.  Finally, we identified multiple overdose 
clusters involving combinations of methadone, commonly used to treat heroin addiction, and the prescription drug Xanax between 2008 and 2012.  We observed dramatic reductions in these clusters after 2012, corresponding to the passage of the 
Methadone Death and Incident Review Act, which increased state oversight of methadone clinics and prescribing physicians.

\section{Conclusions}
\label{sec:conclusions}

Taken together, these two case studies suggest high potential utility 
for prospective surveillance of overdose data using subset scan methods.  
Detected clusters of overdoses which are geographically, 
demographically, and/or behaviorally similar can suggest details of the 
underlying process causing these deaths, and can help local public 
health agencies to target their prevention and response efforts to 
specific neighborhoods and subpopulations.  These efforts can include 
outreach to increase the proportion of opioid addicts receiving 
Medication-Assisted Treatment such as methadone and suboxone, as well as 
patients' compliance with their treatment regimes. Additionally, 
increased supplies of the overdose treatment drug naloxone can be 
directed to potential overdose victims and their local communities.  In 
addition to facilitating targeted and effective health interventions by 
prospective surveillance, these approaches can also be used on 
retrospective data to better understand the effects of drug 
legislation and other policy changes.  Our ongoing work will integrate 
these two methods to better handle multidimensional, correlated 
space-time data, and apply them to new overdose-related data sources, 
including data from the state of Kansas's prescription drug monitoring 
program, in collaboration with public health partners.

\section{Acknowledgments}

A portion of this work was performed while DN and WH were respectively 
visiting professor and visiting researcher at NYU's Center for Urban 
Science and Progress.  We gratefully acknowledge funding support from 
the National Science Foundation (IIS-0953330 and GRFP DGE-1252522) and 
NCSU Laboratory for Analytical Sciences.  The authors wish to thank Eric 
Hulsey (Allegheny County Department of Human Services) for feedback on 
discovered clusters.

\nocite{*}
\bibliographystyle{abbrv}
\bibliography{D4GX}

\end{document}